# Enhancing Understanding of Driving Attributes through Quantitative Assessment of Driver Cognition

Pallabjyoti Kakoti[1a], Mukesh Kumar Kamti[2b], Rauf Iqbal[2c], Eeshankur Saikia[1d]

[1] Department of Applied Sciences, Gauhati University, Guwahati 781014, India

[2] Department of Ergonomics and Human Factors Engineering, National Institute of Industrial Engineering, Mumbai 400087, India

**Corresponding Author:** Eeshankur Saikia[1d]

**Email:** kakotipallab86@gmail.com [a]

mukeshkumar.kamti.2019@nitie.ac.in [b]

raufiqbal@nitie.ac.in [c]

eeshankur@gauhati.ac.in [d]

**Abstract:**

This paper presents a novel approach for analysing EEG data from drivers in a simulated driving test. We focused on the Hurst exponent, Shannon entropy, and fractal dimension as markers of the nonlinear dynamics of the brain. The results show significant trends: Shannon Entropy and Fractal Dimension exhibit variations during driving condition transitions, whereas the Hurst exponent reflects memory retention portraying learning patterns. These findings suggest that the tools of Non-linear Dynamical (NLD) Theory as indicators of cognitive state and driving memory changes for assessing driver performance, and advancing the understanding of non-linear dynamics of human cognition in the context of driving and beyond. Our study reveals the potential of NLD tools to elucidate brain state and system variances, enabling their integration into current Deep Learning and Machine Learning models. This integration can extend beyond driving applications and be harnessed for cognitive learning, thereby improving overall productivity and accuracy levels.

## 1. Introduction:

Driving is a complex skill that requires many cognitive capabilities. With age and experience, these skills evolve over time. The association between cognitive ability and driving performance is well established (De Raedt & Ponjaert, 2000). Experience has a significant impact on how cognitive capacities develop, resulting in increases in several areas, such as response time and decision-making ability (Nabatilan et al., 2012). Individuals develop a level of familiarity and experience with particular stimuli or tasks via repeated exposure, which results in quicker reaction times (Nelson, Thomas & de Haan, 2012). This is frequently seen in activities such as driving, athletics, or video games, where seasoned players react more quickly than inexperienced players. Because the brain may automate some cognitive functions with experience, reaction times shorten as neural pathways become more effective. However, although the experience can enhance cognitive abilities, age-related changes can also have a significant impact on cognitive function (Depestele et. al 2020). Certain cognitive abilities, such as memory, attention, and executive function, may deteriorate with age (Levy, 1994). To improve cognitive performance and treat age-related cognitive decline and neurodegenerative illnesses, it is crucial to understand the interaction between experience, aging, and cognitive capacities.

EEG analysis has been used to study driving performance in a variety of ways, including measuring drivers' brain activity while they are taking part in a simulated or real driving test to identify changes in brain activity linked to fatigue, distraction, and other factors that can impair driving performance (Schier, 2000, Karthaus, Wascher & Getzmann, 2018, Trejo et al., 2015) The study of brain oscillations is another area of EEG analysis in simulated driving (Lin et al., 2005). Rhythmic patterns of brain activity in many frequency bands, including the alpha, beta, theta, and gamma bands, are represented by neural oscillations. Decreased alpha oscillations may indicate higher alertness, while increased theta oscillations may indicate cognitive strain or workload (Teplan 2002). The complexity, regularity, and scaling characteristics of brain activity can also be shown by measurements obtained from EEG data, such as Hurst exponent, Shannon frequency, and fractal dimension. These measurements can be used in the context of simulated driving to evaluate the degree of cognitive engagement, cognitive load, and effectiveness of information processing.

Nonlinear dynamics is a branch of mathematics that studies systems with complicated behaviours that are sensitive to tiny changes in the initial circumstances. It is concerned with the investigation of nonlinear equations and the behaviour of systems, which cannot be simply explained using Newtonian, Hamiltonian, or other linear models. The brain functions as a nonlinear system (Stam, 2006). This is due to the fact that the brain is made up of billions of neurons that are intricately intertwined. Small changes in the activity of one neuron can have a huge impact on the activity of neighbouring neurons. This can result in behavioural, perceptual, and cognitive alterations (Freeman & Vitiello, 2006). Nonlinear properties are frequently observed in brain activity data, such as electroencephalography (EEG) or functional magnetic resonance imaging (fMRI) (Portnova et al., 2018). Linear models may not be sufficient to capture the underlying dynamics or extract meaningful information from such data.

Hurst exponent, a measure of temporal dynamics' "long-term memory," has also been investigated in recent years as a tool for EEG signal interpretation (Yean, Wen, et al., 2019). The Hurst exponent is a mathematical measure used to analyse the long-term memory and predictability of a time series (Cerqueti, R, and Mattera, 2023). The Hurst exponent is used in EEG signal analysis to measure the fractal scaling characteristics of the signal, which are connected to the long-range correlations and self-similarities of the time series. By examining the complexity of the EEG signal, the Hurst exponent can be used to distinguish between healthy and pathological brain states (Choong et al. 2021). For instance, research has revealed that the Hurst exponents of Alzheimer's patients' EEG signals are lower than those of healthy controls, indicating a loss of complexity and long-range correlations in brain activity (Amezquita-Sanchez et al., 2019). The Hurst exponent has also been found to be altered in various neurological conditions including Parkinson's disease and epilepsy. (Kamalakannan, et al.. 2021) (Abbasi, Muhammad U. et al. 2019)

Entropy is a different approach that can be used to extract regular information from EEG datasets (Kannathal et al. 2005). Entropy is a nonlinear property that measures the degree of randomness in a system. Because it cannot be evaluated by frequency relative power derived from linear analysis, it is a useful tool for classifying mental states based on the degree of temporal and spectral irregularity in the EEG signal (M. Ignaccolo, et. Al. 2010).

The complexity and self-similarity of the EEG data were measured mathematically using the fractal dimension in EEG analysis. It sheds light on the temporal structure and scaling characteristics of the electrical activity of the brain. Fractal dimension is derived from the concept of fractals, which are geometric objects that exhibit self-similarity at different scales. It is calculated by splitting the signal into smaller segments and calculating the fluctuation within each segment. The complexity of the signal increases with fractal dimension.

In EEG analysis, fractal dimension has been used to investigate a number of cognitive processes, such as decision-making, memory, and attention (John et. al 2015). In addition, they have been used for biomedical signal processing (Das et. al 2013) (Das et. al 2014). For instance, research has revealed that EEG signals from people with attention deficit hyperactivity disorder (ADHD) have higher fractal dimensions in their EEG signals than those without ADHD (Dawi et. al 2021). This shows that the increased complexity of brain activity may be linked to ADHD. Additionally, research on brain illnesses, such as Schizophrenia, Depression and Alzheimer's disease, has made use of the fractal dimension (Kakoti, Syiemlieh, & Saikia, 2021). For instance, research has revealed that EEG signals from individuals with Alzheimer's disease are less fractal than those from individuals without Alzheimer's disease (Ahmadlou et. al. 2011). This suggests that Alzheimer's disease may be linked to a decline in cognitive complexity.

In this study, we investigated the relationship between age, driving experience, and cognitive processes in driving performance by analysing EEG data collected from drivers and students during a simulated driving test and an on-road experiment. Our approach focused on utilizing nonlinear dynamical parameters to unveil the state of cognition and driving experience embedded in the EEG data. Similar analyses were conducted for EEG & fMRI brain signals (Kakoti, Syiemlieh, & Saikia, 2021), solar-cycle activities (Syiemlieh et al., 2022), and also nano-material fabrication (Saha et al., 2020). Specifically, we computed the Hurst exponent to reveal the retention of memory and experience, while the Fractal Dimension and Shannon Entropy shed light on the complexity of neuronal firing and ease of cognitive processes. By leveraging these nonlinear parameters, we gained insights into the effects of age, driving experience, and task difficulty on driving ability.

## 2. Materials & Methods:

Ten male three-wheeler drivers were selected for this cross-sectional study from five distinct three-wheeler stands (Autorickshaw stands) in Mumbai, India. Participants in the study must possess a valid driving license, have at least two years of driving experience, and log at least five hours of driving each day.

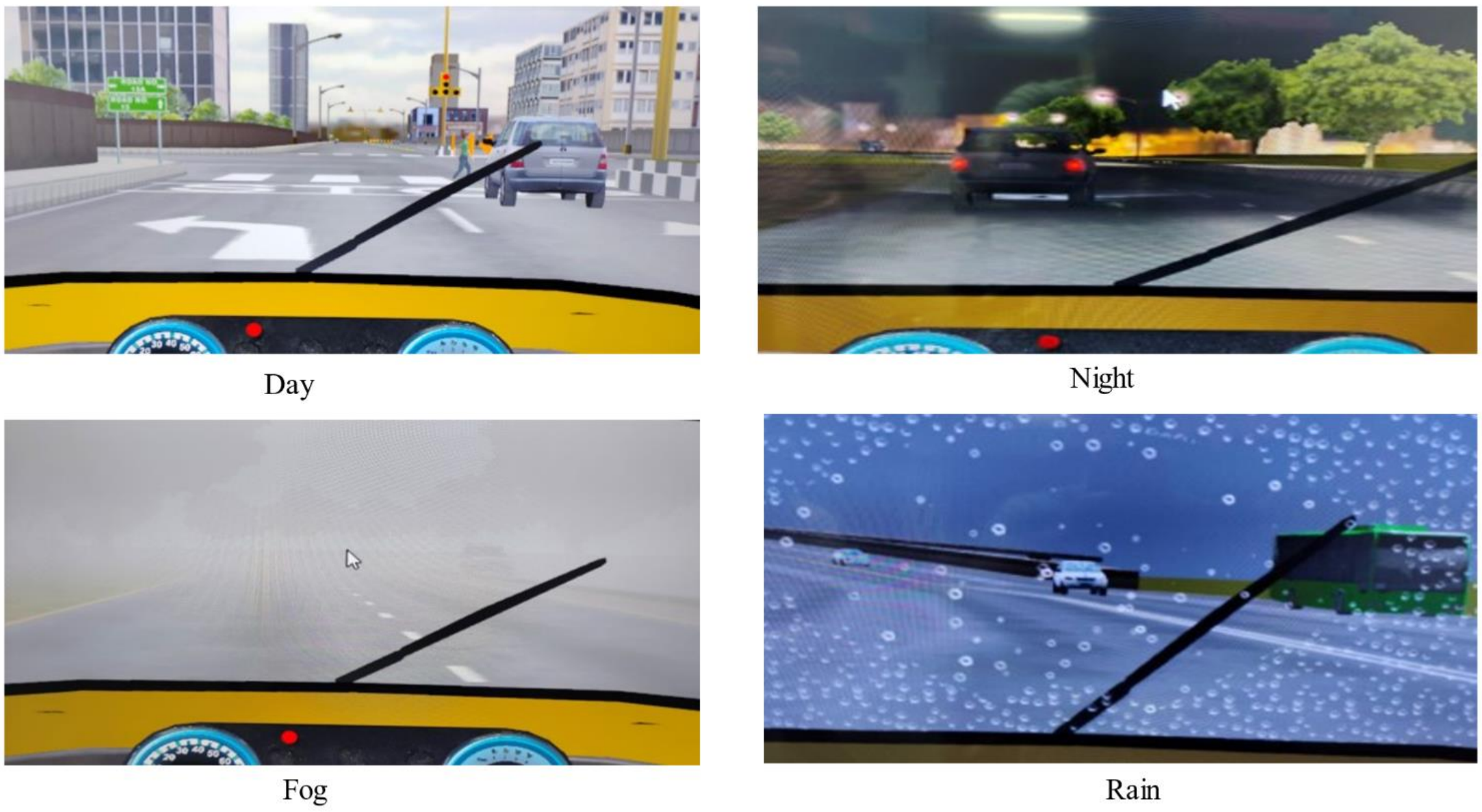

*Figure 1: Four Environment conditions for the driving simulator*

The experiment was conducted on drivers in the Ergonomics Laboratory of the National Institute of Industrial Engineering (NITIE), Mumbai, using a fixed three-wheeler driving simulator (Technotrov Systems Pvt. Ltd., Maharashtra, Mumbai; see Fig. 1). This driving simulator is an exact reproduction of a three-wheeled vehicle, complete with all its characteristics. For this experiment, four different traffic situations were selected: low traffic in cities, high traffic in cities, and low traffic on highways. All subjects had 3 min of practice time to become accustomed to the system and procedure, followed by 15 min of EEG recording.

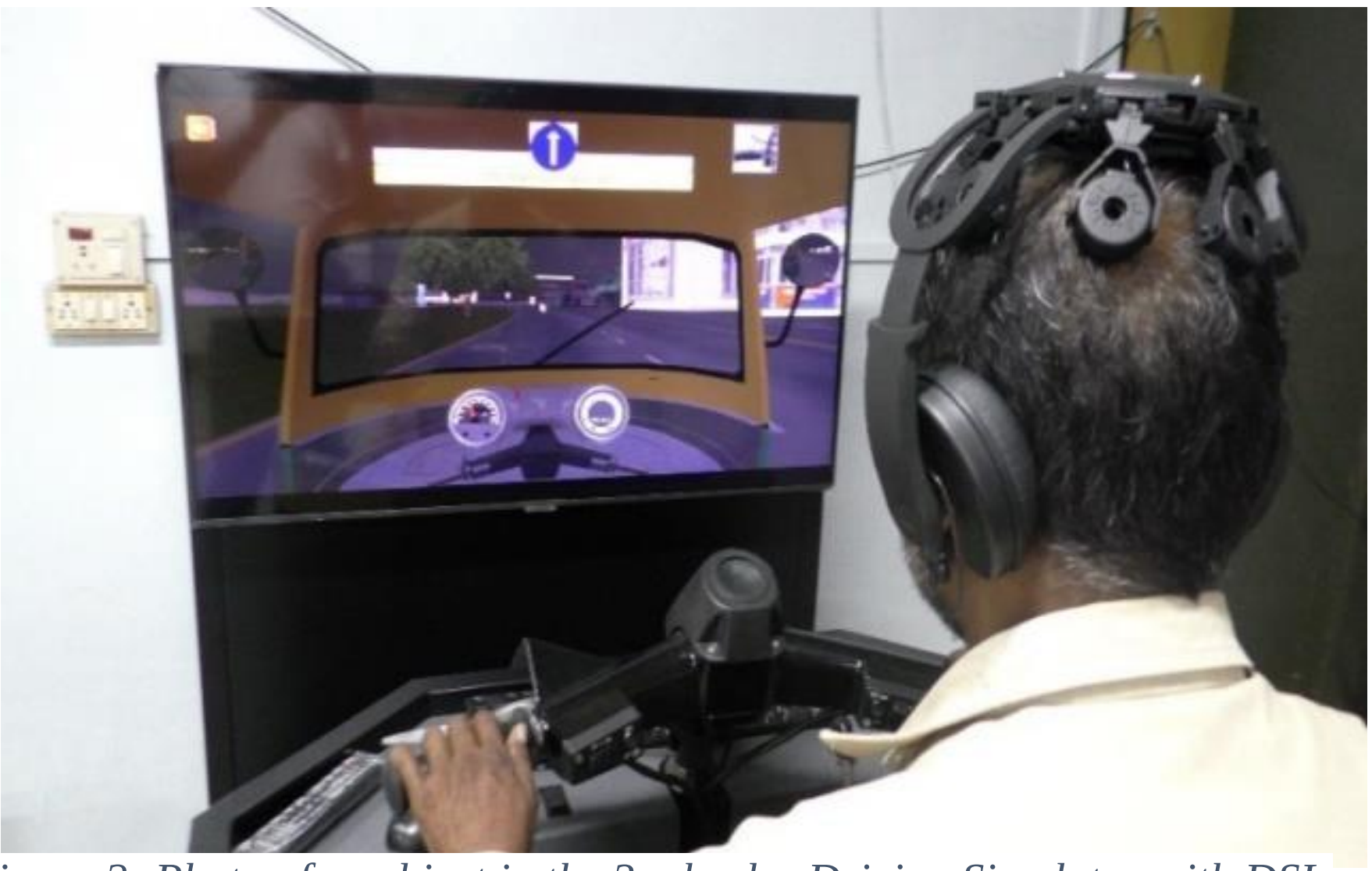

*Figure 2: Photo of a subject in the 3-wheeler Driving Simulator with DSI-7 Wireless EEG Headset mounted on head.*

The experiments consisted of four EEG sessions, each lasting approximately four hours, which included four different environments and four traffic conditions for each subject. Thirteen 3-wheeler driver EEG data were collected on the road during high and low traffic conditions for 5 min. Wearable Sensing DSI-7 EEG Headset (Figure 2) was used to collect data from 20 subjects, which is a research-grade EEG sensing device with 8 dry-application sensors, including one for reference (LE) and seven for recording brain wave activity (F3, F4, C3, C4, P3, Pz, P4) (Matthews et. al., 2007). The sampling frequency was set to 300 Hz for the EEG data streamer. The data files that were used for the analyses reported below for all experiments can be found at https://osf.io/8ksyq.

## 2.2 Data Pre-Processing

EEGLab (Delorme & Makeig, 2004) in MATLAB was used to perform EEG pre-processing using the PREP pipeline (Bigdely et. al., 2015). Detrending was carried out to eliminate any cyclical or other patterns and calibrate the thresholds. To eliminate linear trends, a 1 Hz high pass filter was applied. A 50 Hz notch filter was used to eliminate the harsh spectral peaks at 50 Hz. For subsequent processing and feature extraction, the F4 electrode was selected for all individuals based on spectral decomposition and signal quality.

### 2.3 Estimation of Hurst Exponent: Nonlinear Dynamical Features

The Hurst Exponent, which depends on the power law (Suyal et al., 2009), was calculated using the Rescaled Range method (R/S) (Mandelbrot and Wallis, 1968).

$$(R/S)_w = kw^H \tag{1}$$

Where S is the standard deviation of the independent variable $x_i$ within the window w, k is a constant, w is the breadth of the temporal window, and

$$|S(t_0, w)|^2 = \frac{1}{w-1} \sum_{i=t_0}^{t_0+w+1} \left[x_i - \underline{x}(t_0, w)\right]^2 \tag{2}$$

With average as,

$$\underline{x}(t_0, w) = \frac{1}{w} \sum_{i=t_0}^{t_0+w+1} x_i \tag{3}$$

And R, the range in the time-series, defined as

$$R(t_0, w) = max_1 \leq i \leq wy_i(t_0, w) - min_1 \leq i \leq wy_i(t_0, w) \tag{4}$$

With the new variables $y_i, i = 1,2,3, \ldots, w$ as,

$$y_i(t_0, w) = \sum_{k=t_0}^{t_0+i+1} [x_k - \bar{x}(t_0, w)] \tag{5}$$

(R/S) was calculated for various time instances, averaged for epochs, and plotted against a log-log axis. The value of H, which ranges from 0 to 1, was determined using the linear regression slope. A time series with a value of H = 0.5 exhibits pure random walking or Brownian motion. On the other hand, H between 0.5 and 1.0 indicates a stable time series. A higher H indicates that the time series has a longer memory and a higher long-term positive autocorrelation or more frequent or persistent deviations. H between 0 and 0.5 indicates anti-persistence, whereas H is more or less equal to 0.5, indicating a random time series (Suyal et al., 2009). The Hurst exponent was calculated for each of the subjects' 16 sessions over the 10 participants' EEG time series.

## 2.4 Estimation of Fractal Dimension

Fractal dimension is a mathematical notion that measures the degree to which a self-similar entity occupies space and is used to measure the complexity of a self-similar object. The box-counting method is a typical method for estimating an object's fractal dimension. It consists of covering the element with a grid of boxes and counting the number of boxes containing a portion of the element. The following is the relationship between the number of boxes $N$ and the box size $r$:

$$N \sim r^{(-D)} \tag{6}$$

where the symbol "~" means "proportional to." Taking the logarithm of both sides of the equation yields

$$\log N \sim -D \log r \tag{7}$$

The slope of the line obtained by plotting $\log N$ against $\log r$ gives an estimate of the fractal dimension $D$.

Consider a Koch curve, which is a fractal structure created by constantly adding smaller equilateral triangles to each side of an original triangle. The fractal dimension of the Koch curve was approximately 1.26. The fractal dimension can be estimated using the box-counting approach by covering the curve with a grid of boxes and counting the number of boxes that contain a part of the curve for different box sizes. Using the above equation, the relationship between the number of boxes and the box size can then be used to estimate the fractal dimension.

### 2.4 Estimation of Shannon Entropy

Because brain activity is a highly dynamic and complicated process involving the interplay of numerous separate neural networks functioning at various frequencies and with differing degrees of synchronization, EEG signals are non-linear, non-stationary, and random. As a result, numerous methods for nonlinear analysis, including entropy, have been proposed to effectively capture the randomness of nonlinear time series data (Phung et al., 2014).

Let X be a set of finite discrete random variables $\mathrm{X} = \{x_1, x_2, x_3, \ldots, x_m\}$, then Shannon entropy, $\mathrm{S(X)}$, is defined as:

$$\mathrm{S(X)} = -\mathrm{c} \sum_{i=0}^{m} p(x_i) \ln p(x_i) \tag{8}$$

where $\mathrm{c}$ is a positive constant acting as a measuring unit and $p(x_i)$ is probability of $x_i \in X$, satisfying:

$$\sum_{i=0}^{m} p(x_i) = 1 \tag{9}$$

In general, more entropy denotes chaotic or more complex systems, and hence less predictability.

## 3. <u>Results & Conclusion:</u>

The values for H, D & S for all the subjects are summarized in the Table 1 and Pearson Correlation Matrix between all the variables are plotted in Fig 3 below.

We conducted an F-Test Two-Sample for Variances to examine the differences in variances between the variables under investigation. The p-values obtained for all pairwise comparisons of variables were found to be less than 0.05 as shown in Table 2, indicating significant differences in variances between the variables.

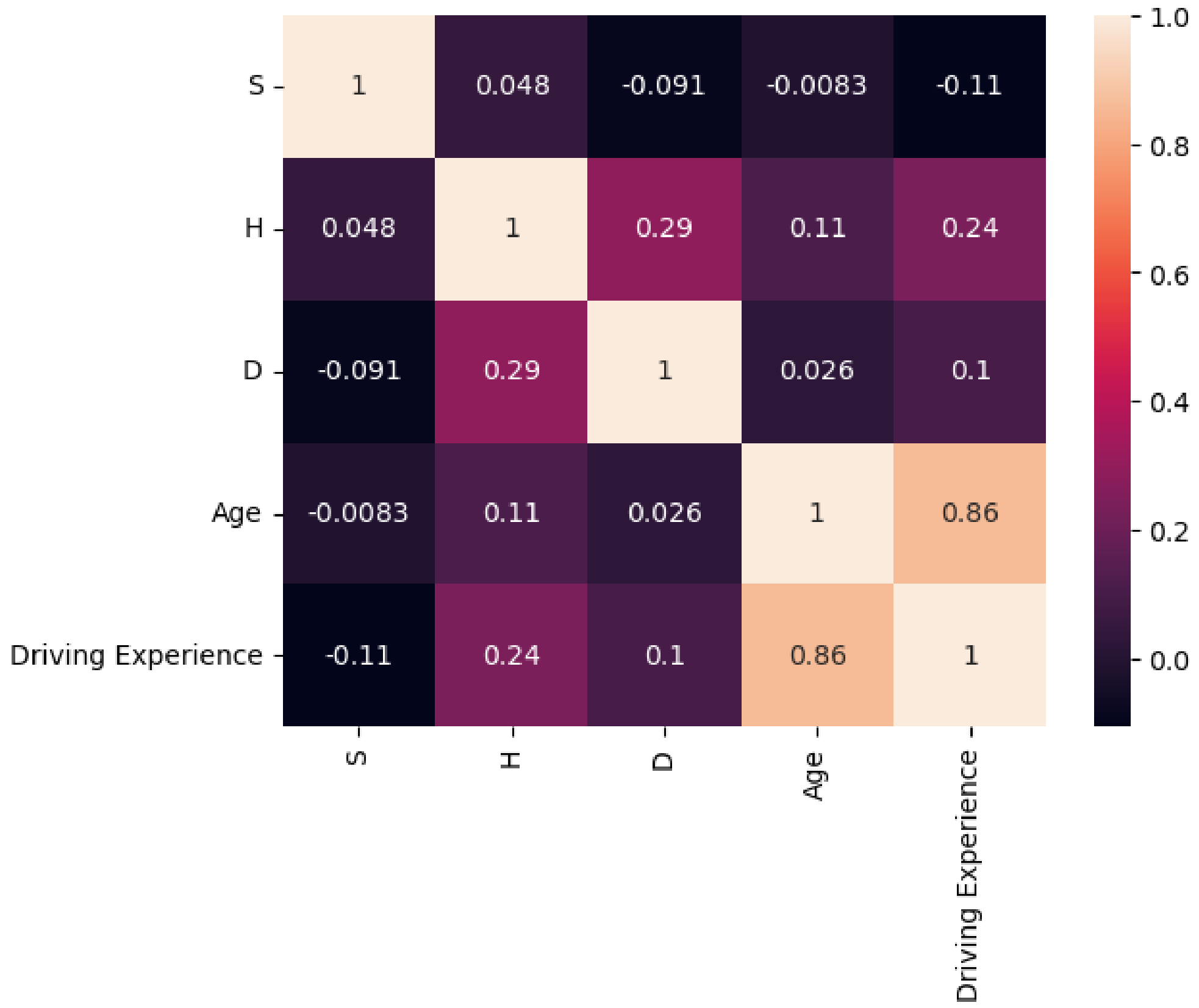

*Figure 3: Pearson correlation matrix for H, D, S, Driving Experience and Age, clearly showing importance of incorporating H, D, S while understanding and quantifying Driving Experience.*

When the difficulty level or weather changed, the driving simulator data analysis showed significant fluctuations in Shannon Entropy (S) and Fractal Dimension (D) as visible in Fig 4 and Fig 5. While D, which denotes complexity or variations in neural patterns, rose during changes in driving circumstances or sessions, S, which stands for chaos and unpredictability, showed considerable swings. Thus, S and D are appropriate markers for investigating sudden changes in driving-related mental states which may be found in the Pearson Correlation Matrix as shown in Fig 3.

Investigations were also conducted on the connection between driving experience and Hurst Exponent (H). H levels varied across driving sessions in subjects of older age and with little prior driving experience as shown in Fig 6. Given that the Hurst Exponent is used to analyze repetitive or memory-related patterns in EEG, the observed fluctuations in H showed weaker retention of driving memory. This research implies that those who have little driving experience and are older may have trouble remembering information about driving over the course of several sessions.

The study also found that H increased when comparing the first and last sessions for each subject, except for those with low driving experience. This suggests that H can serve as a marker for memory retention. Participants with higher H values between the initial and final sessions showed better retention of driving-related information, indicating the role of H in assessing memory retention during driving.

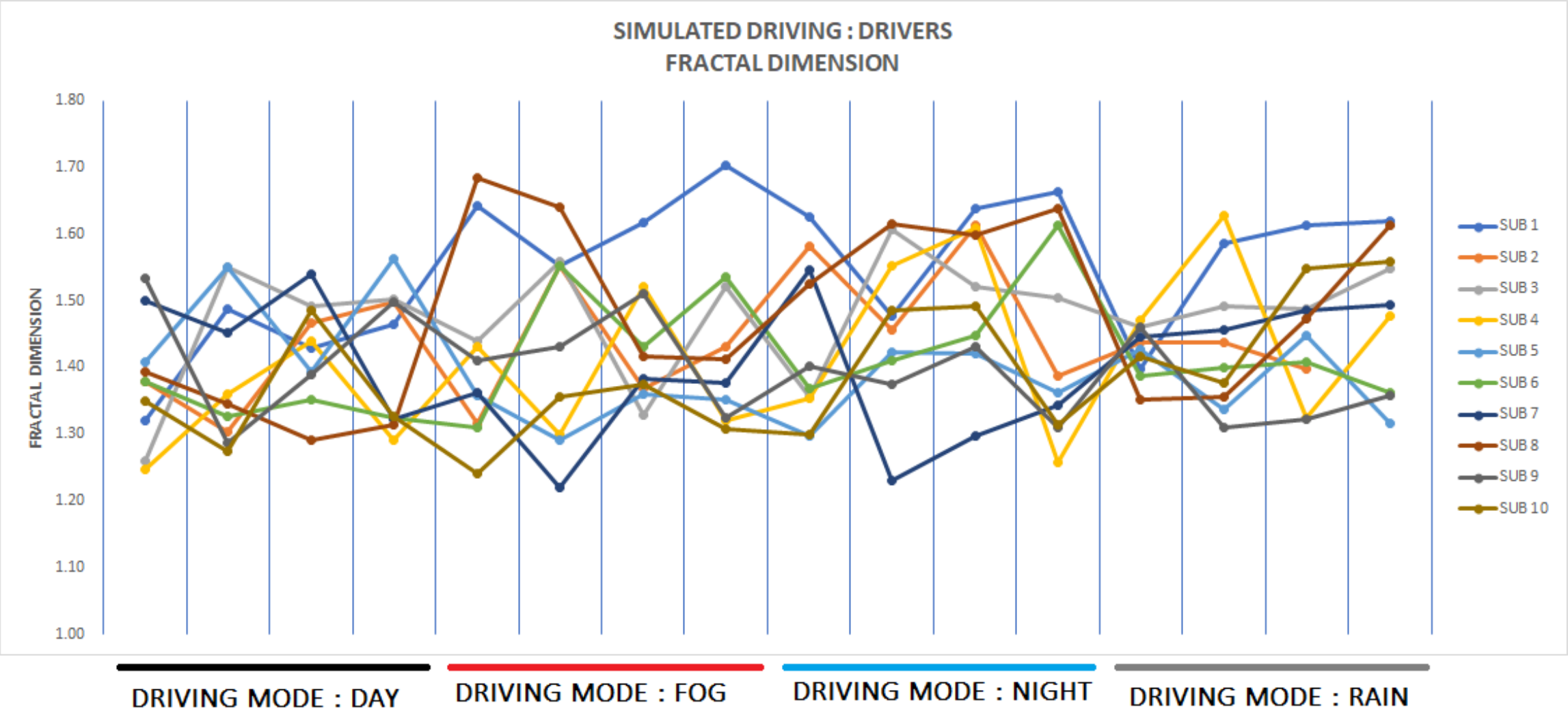

*Figure 4: Graph showing the variation of Fractal Dimension in Simulated Driving Experiment for 10 subjects. Each vertical gridline represents the end of an EEG session.*

The study findings suggest that S, D, and H can serve as indicators of changes in the mental state and driving memory during driving.

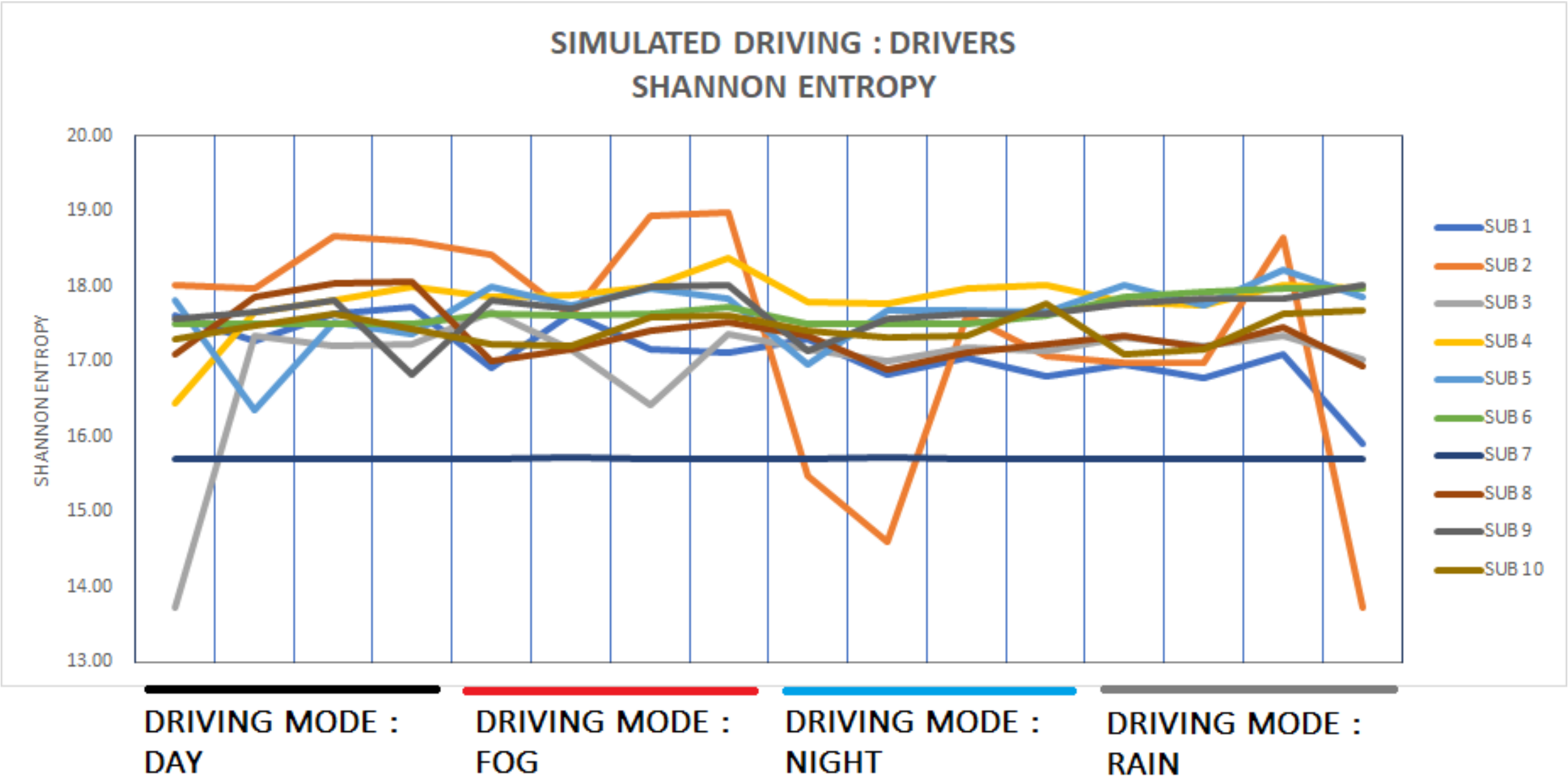

*Figure 5: Graph showing the variation of Shannon Entropy in Simulated Driving Experiment for 10 subjects. Each vertical gridline represents the end of an EEG session.*

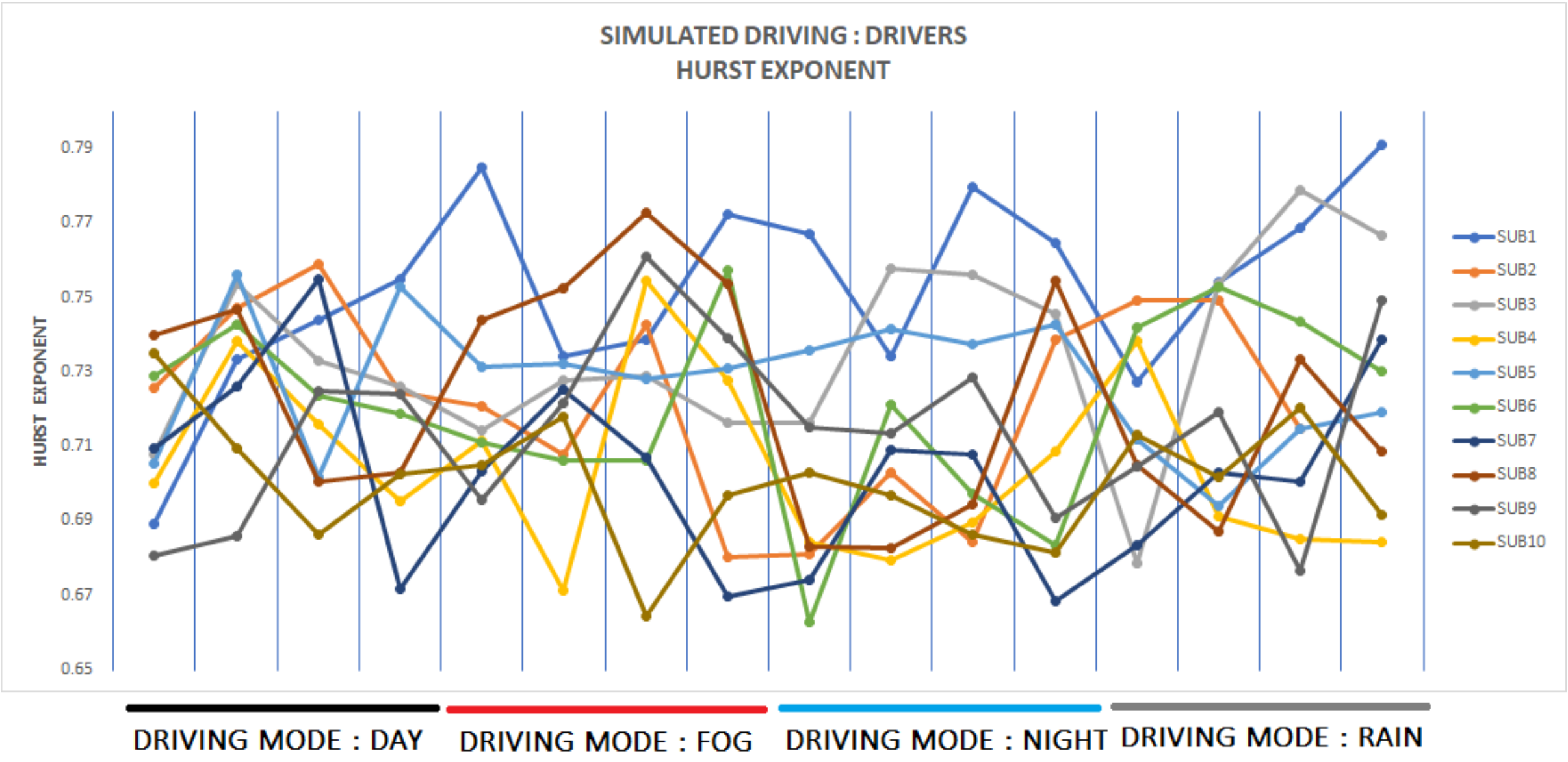

*Figure 6: Graph showing the variation of Hurst Exponent in Simulated Driving Experiment for 10 subjects. Each vertical gridline represents the end of an EEG session.*

These measures have the potential to develop new approaches for assessing driver performance and identifying safety risks. Moreover, this study demonstrates the applicability of nonlinear dynamical statistics to explain the differences among systems. By applying nonlinear dynamical statistics to the analysis of drivers, this study reveals the relationship between driving experience and age. Therefore, incorporating nonlinear dynamical statistics into existing Deep Learning and Machine Learning models may enhance overall learning and accuracy. Additional research is required to investigate the optimal integration of these markers and validate their efficacy using larger and more diverse datasets derived from cognitive learning experiments.

## CRediT Authorship Contribution Statement:

**Pallabjyoti Kakoti:** Formal analysis, Software, Visualization, Writing – original draft. **Mukesh Kamti Debnath:** Data curation, Investigation, Methodology **Rauf Iqbal:** Project administration, Funding acquisition, Resources **Eeshankur Saikia:** Conceptualization, Supervision, Writing – review & editing

## Acknowledgment:

MKD and RI are thankful towards the group of autorickshaw drivers for giving consent for participating in this study and also express gratitude towards Ergonomics Lab, NITIE for funding this study for setting up the laboratory, and providing remuneration to the participants of the study. PK and ES are thankful to Ergonomics Lab, NITIE for facilitating the academic collaboration.

## Conflict of Interest Statement:

The authors declare no conflicts of interest.

## Data Availability Statement:

The EEG data that has been used for the analysis can be found on the following URL: https://osf.io/8ksyq

## Compliance with Ethical Standards:

All procedures performed in the study involved human participants and all procedures has been carried out in accordance with The Code of Ethics of the World Medical Association (Declaration of Helsinki) for experiments involving humans.

## Informed Consent:

Informed consent was obtained from all individual participants included in the study.

## Tables:

***Table 1***: *Tabular data showing the variation in Shannon Entropy (S), Hurst Exponent (H) and Fractal Dimension (D) for the subjects for different sessions and difficulty levels. Sessions were divided into 4 sets comprising of different weather and traffic modes (Highway Low & High Traffic, City Low & High Traffic). Age and Driving Experience of the subjects were also recorded for review.*

| Subject | Session | S | H | D | Age | Driving Experience |
|---|---|---|---|---|---|---|
| 1 | DAY_DRIVING_A | 17.62 | 0.69 | 1.32 | 44 | 24 |
| | DAY_DRIVING_B | 17.27 | 0.73 | 1.49 | | |
| | DAY_DRIVING_C | 17.65 | 0.74 | 1.43 | | |
| | DAY_DRIVING_D | 17.72 | 0.75 | 1.46 | | |
| | FOG_DRIVING_A | 16.91 | 0.78 | 1.64 | | |
| | FOG_DRIVING_B | 17.63 | 0.73 | 1.55 | | |
| | FOG_DRIVING_C | 17.16 | 0.74 | 1.62 | | |
| | FOG_DRIVING_D | 17.11 | 0.77 | 1.70 | | |
| | NIGHT_DRIVING_A | 17.30 | 0.77 | 1.63 | | |
| | NIGHT_DRIVING_B | 16.83 | 0.73 | 1.48 | | |
| | NIGHT_DRIVING_C | 17.04 | 0.78 | 1.64 | | |
| | NIGHT_DRIVING_D | 16.80 | 0.76 | 1.66 | | |
| | RAIN_DRIVING_A | 16.97 | 0.73 | 1.40 | | |
| | RAIN_DRIVING_B | 16.79 | 0.75 | 1.59 | | |
| | RAIN_DRIVING_C | 17.09 | 0.77 | 1.61 | | |
| | RAIN_DRIVING_D | 15.91 | 0.79 | 1.62 | | |
| 2 | DAY_DRIVING_A | 18.02 | 0.73 | 1.38 | 53 | 17 |
| | DAY_DRIVING_B | 17.98 | 0.75 | 1.30 | | |
| | DAY_DRIVING_C | 18.68 | 0.76 | 1.47 | | |
| | DAY_DRIVING_D | 18.59 | 0.72 | 1.50 | | |
| | FOG_DRIVING_A | 18.41 | 0.72 | 1.32 | | |
| | FOG_DRIVING_B | 17.61 | 0.71 | 1.55 | | |
| | FOG_DRIVING_D | 18.94 | 0.74 | 1.37 | | |
| | NIGHT_DRIVING_A | 18.98 | 0.68 | 1.43 | | |
| | NIGHT_DRIVING_B | 15.48 | 0.68 | 1.58 | | |
| | NIGHT_DRIVING_C | 14.61 | 0.70 | 1.46 | | |
| | NIGHT_DRIVING_D | 17.62 | 0.68 | 1.61 | | |
| | RAIN_DRIVING_A | 17.08 | 0.74 | 1.39 | | |
| | RAIN_DRIVING_B | 16.99 | 0.75 | 1.44 | | |
| | RAIN_DRIVING_C | 16.99 | 0.75 | 1.44 | | |
| | RAIN_DRIVING_D | 18.65 | 0.71 | 1.40 | | |
| 3 | DAY_DRIVING_A | 13.73 | 0.71 | 1.26 | 39 | 21 |
| | DAY_DRIVING_B | 17.35 | 0.75 | 1.55 | | |
| | DAY_DRIVING_C | 17.20 | 0.73 | 1.49 | | |
| | DAY_DRIVING_D | 17.23 | 0.73 | 1.50 | | |
| | FOG_DRIVING_A | 17.66 | 0.71 | 1.44 | | |
| | FOG_DRIVING_B | 17.17 | 0.73 | 1.56 | | |
| | FOG_DRIVING_C | 16.42 | 0.73 | 1.33 | | |
| | FOG_DRIVING_D | 17.37 | 0.72 | 1.52 | | |
| | NIGHT_DRIVING_A | 17.17 | 0.72 | 1.36 | | |
| | NIGHT_DRIVING_B | 17.01 | 0.76 | 1.61 | | |
| | NIGHT_DRIVING_C | 17.20 | 0.76 | 1.52 | | |
| | NIGHT_DRIVING_D | 17.13 | 0.75 | 1.50 | | |
| | RAIN_DRIVING_A | 17.31 | 0.68 | 1.46 | | |
| | RAIN_DRIVING_B | 17.21 | 0.75 | 1.49 | | |
| | RAIN_DRIVING_C | 17.35 | 0.78 | 1.49 | | |
| | RAIN_DRIVING_D | 17.03 | 0.77 | 1.55 | | |
| 4 | DAY_DRIVING_A | 16.45 | 0.70 | 1.25 | 27 | 02 |
| | DAY_DRIVING_B | 17.66 | 0.74 | 1.36 | | |
| | DAY_DRIVING_C | 17.81 | 0.72 | 1.44 | | |
| | DAY_DRIVING_D | 18.00 | 0.70 | 1.29 | | |
| | FOG_DRIVING_A | 17.86 | 0.71 | 1.43 | | |
| | FOG_DRIVING_B | 17.88 | 0.67 | 1.30 | | |
| | FOG_DRIVING_C | 17.99 | 0.75 | 1.52 | | |
| | FOG_DRIVING_D | 18.37 | 0.73 | 1.32 | | |
| | NIGHT_DRIVING_A | 17.80 | 0.68 | 1.35 | | |
| | NIGHT_DRIVING_B | 17.77 | 0.68 | 1.55 | | |
| | NIGHT_DRIVING_C | 17.98 | 0.69 | 1.61 | | |
| | NIGHT_DRIVING_D | 18.01 | 0.71 | 1.26 | | |
| | RAIN_DRIVING_A | 17.80 | 0.74 | 1.47 | | |
| | RAIN_DRIVING_B | 17.74 | 0.69 | 1.63 | | |
| | RAIN_DRIVING_C | 18.02 | 0.69 | 1.33 | | |
| | RAIN_DRIVING_D | 17.98 | 0.68 | 1.48 | | |
| 5 | DAY_DRIVING_A | 17.81 | 0.71 | 1.41 | 45 | 18 |
| | DAY_DRIVING_B | 16.36 | 0.76 | 1.55 | | |
| | DAY_DRIVING_C | 17.52 | 0.70 | 1.39 | | |
| | DAY_DRIVING_D | 17.36 | 0.75 | 1.56 | | |
| | FOG_DRIVING_A | 17.99 | 0.73 | 1.36 | | |
| | FOG_DRIVING_B | 17.74 | 0.73 | 1.29 | | |
| | FOG_DRIVING_C | 17.97 | 0.73 | 1.36 | | |
| | FOG_DRIVING_D | 17.83 | 0.73 | 1.35 | | |
| | NIGHT_DRIVING_A | 16.97 | 0.74 | 1.30 | | |
| | NIGHT_DRIVING_B | 17.68 | 0.74 | 1.42 | | |
| | NIGHT_DRIVING_C | 17.69 | 0.74 | 1.42 | | |
| | NIGHT_DRIVING_D | 17.66 | 0.74 | 1.36 | | |
| | RAIN_DRIVING_A | 18.02 | 0.71 | 1.43 | | |
| | RAIN_DRIVING_B | 17.74 | 0.69 | 1.34 | | |
| | RAIN_DRIVING_C | 18.22 | 0.71 | 1.45 | | |
| | RAIN_DRIVING_D | 17.85 | 0.72 | 1.32 | | |
| 6 | DAY_DRIVING_A | 17.51 | 0.73 | 1.38 | 63 | 41 |
| | DAY_DRIVING_B | 17.50 | 0.74 | 1.33 | | |
| | DAY_DRIVING_C | 17.50 | 0.72 | 1.35 | | |
| | DAY_DRIVING_D | 17.51 | 0.72 | 1.32 | | |
| | FOG_DRIVING_A | 17.63 | 0.71 | 1.31 | | |
| | FOG_DRIVING_B | 17.61 | 0.71 | 1.55 | | |
| | FOG_DRIVING_C | 17.64 | 0.71 | 1.43 | | |
| | FOG_DRIVING_D | 17.72 | 0.76 | 1.54 | | |
| | NIGHT_DRIVING_A | 17.51 | 0.66 | 1.37 | | |
| | NIGHT_DRIVING_B | 17.49 | 0.72 | 1.41 | | |
| | NIGHT_DRIVING_C | 17.51 | 0.70 | 1.45 | | |
| | NIGHT_DRIVING_D | 17.62 | 0.68 | 1.61 | | |
| | RAIN_DRIVING_A | 17.86 | 0.74 | 1.39 | | |
| | RAIN_DRIVING_B | 17.93 | 0.75 | 1.40 | | |
| | RAIN_DRIVING_C | 17.96 | 0.74 | 1.41 | | |
| | RAIN_DRIVING_D | 17.96 | 0.73 | 1.36 | | |
| 7 | DAY_DRIVING_A | 15.71 | 0.71 | 1.50 | 42 | 18 |
| | DAY_DRIVING_B | 15.71 | 0.73 | 1.45 | | |
| | DAY_DRIVING_C | 15.70 | 0.75 | 1.54 | | |
| | DAY_DRIVING_D | 15.71 | 0.67 | 1.32 | | |
| | FOG_DRIVING_A | 15.71 | 0.70 | 1.36 | | |
| | FOG_DRIVING_B | 15.71 | 0.73 | 1.22 | | |
| | FOG_DRIVING_C | 15.71 | 0.71 | 1.38 | | |
| | FOG_DRIVING_D | 15.71 | 0.67 | 1.38 | | |
| | NIGHT_DRIVING_A | 15.69 | 0.67 | 1.55 | | |
| | NIGHT_DRIVING_B | 15.71 | 0.71 | 1.23 | | |
| | NIGHT_DRIVING_C | 15.71 | 0.71 | 1.30 | | |
| | NIGHT_DRIVING_D | 15.71 | 0.67 | 1.34 | | |
| | RAIN_DRIVING_A | 15.70 | 0.68 | 1.45 | | |
| | RAIN_DRIVING_B | 15.71 | 0.70 | 1.46 | | |
| | RAIN_DRIVING_C | 15.70 | 0.70 | 1.49 | | |
| | RAIN_DRIVING_D | 15.70 | 0.74 | 1.49 | | |
| 8 | DAY_DRIVING_A | 17.09 | 0.74 | 1.39 | 39 | 10 |
| | DAY_DRIVING_B | 17.85 | 0.75 | 1.35 | | |
| | DAY_DRIVING_C | 18.03 | 0.70 | 1.29 | | |
| | DAY_DRIVING_D | 18.05 | 0.70 | 1.31 | | |
| | FOG_DRIVING_A | 17.01 | 0.74 | 1.68 | | |
| | FOG_DRIVING_B | 17.17 | 0.75 | 1.64 | | |
| | FOG_DRIVING_C | 17.40 | 0.77 | 1.42 | | |
| | FOG_DRIVING_D | 17.52 | 0.75 | 1.41 | | |
| | NIGHT_DRIVING_A | 17.35 | 0.68 | 1.53 | | |
| | NIGHT_DRIVING_B | 16.89 | 0.68 | 1.62 | | |
| | NIGHT_DRIVING_C | 17.12 | 0.69 | 1.60 | | |
| | NIGHT_DRIVING_D | 17.22 | 0.75 | 1.64 | | |
| | RAIN_DRIVING_A | 17.34 | 0.70 | 1.35 | | |
| | RAIN_DRIVING_B | 17.18 | 0.69 | 1.36 | | |
| | RAIN_DRIVING_C | 17.44 | 0.73 | 1.47 | | |
| | RAIN_DRIVING_D | 16.95 | 0.71 | 1.61 | | |
| 9 | DAY_DRIVING_A | 17.56 | 0.68 | 1.53 | 27 | 05 |
| | DAY_DRIVING_B | 17.65 | 0.69 | 1.29 | | |
| | DAY_DRIVING_C | 17.81 | 0.72 | 1.39 | | |
| | DAY_DRIVING_D | 16.83 | 0.72 | 1.50 | | |
| | FOG_DRIVING_A | 17.82 | 0.70 | 1.41 | | |
| | FOG_DRIVING_B | 17.71 | 0.72 | 1.43 | | |
| | FOG_DRIVING_C | 17.99 | 0.76 | 1.51 | | |
| | FOG_DRIVING_D | 18.01 | 0.74 | 1.33 | | |
| | NIGHT_DRIVING_A | 17.13 | 0.72 | 1.40 | | |
| | NIGHT_DRIVING_B | 17.57 | 0.71 | 1.38 | | |
| | NIGHT_DRIVING_C | 17.63 | 0.73 | 1.43 | | |
| | NIGHT_DRIVING_D | 17.62 | 0.69 | 1.31 | | |
| | RAIN_DRIVING_A | 17.77 | 0.70 | 1.46 | | |
| | RAIN_DRIVING_B | 17.84 | 0.72 | 1.31 | | |
| | RAIN_DRIVING_C | 17.84 | 0.68 | 1.32 | | |
| | RAIN_DRIVING_D | 18.01 | 0.75 | 1.36 | | |
| 10 | DAY_DRIVING_A | 17.30 | 0.73 | 1.35 | 44 | 10 |
| | DAY_DRIVING_B | 17.48 | 0.71 | 1.28 | | |
| | DAY_DRIVING_C | 17.63 | 0.69 | 1.49 | | |
| | DAY_DRIVING_D | 17.44 | 0.70 | 1.33 | | |
| | FOG_DRIVING_A | 17.23 | 0.70 | 1.24 | | |
| | FOG_DRIVING_B | 17.21 | 0.72 | 1.35 | | |
| | FOG_DRIVING_C | 17.60 | 0.66 | 1.38 | | |
| | FOG_DRIVING_D | 17.61 | 0.70 | 1.31 | | |
| | NIGHT_DRIVING_A | 17.41 | 0.70 | 1.30 | | |
| | NIGHT_DRIVING_B | 17.33 | 0.70 | 1.49 | | |
| | NIGHT_DRIVING_C | 17.35 | 0.69 | 1.49 | | |
| | NIGHT_DRIVING_D | 17.77 | 0.68 | 1.31 | | |
| | RAIN_DRIVING_A | 17.10 | 0.71 | 1.42 | | |
| | RAIN_DRIVING_B | 17.17 | 0.70 | 1.38 | | |
| | RAIN_DRIVING_C | 17.63 | 0.72 | 1.55 | | |
| | RAIN_DRIVING_D | 17.67 | 0.69 | 1.56 | | |

***Table 2**: Tabular data showing the F-test Two-Sample for checking the variance between the independent variables (Subject, Simulator Traffic conditions and Environment conditions) and dependent variables (Shannon Entropy (S), Hurst Exponent (H) and Fractal Dimension (D). he p-values obtained for all pairwise comparisons of variables were found to be less than 0.05, indicating significant differences in variances between the variables.*

| | *Subject* | *S* |
|---|---|---|
| Mean | 5.522012579 | 17.29931 |
| Variance | 8.276411114 | 0.652137 |
| Observations | 159 | 159 |
| df | 158 | 158 |
| F | 12.69121809 | |
| P(F<=f) one-tail | 5.62503E-47 | |
| F Critical one-tail | 1.300182044 | |

| | *Subject* | *H* |
|---|---|---|
| Mean | 5.52201258 | 0.719811 |
| Variance | 8.27641111 | 0.000798 |
| Observations | 159 | 159 |
| df | 158 | 158 |
| F | 10370.592 | |
| P(F<=f) one-tail | 6.436E-272 | |
| F Critical one-tail | 1.30018204 | |

| | *Subject* | *D* |
|---|---|---|
| Mean | 5.5220126 | 1.436101 |
| Variance | 8.2764111 | 0.012073 |
| Observations | 159 | 159 |
| df | 158 | 158 |
| F | 685.51322 | |
| P(F<=f) one-tail | 8.31E-179 | |
| F Critical one-tail | 1.300182 | |

| | *Environment* | *S* |
|---|---|---|
| Mean | 2.503144654 | 17.29931 |
| Variance | 1.264230555 | 0.652137 |
| Observations | 159 | 159 |
| df | 158 | 158 |
| F | 1.938596991 | |
| P(F<=f) one-tail | 1.92043E-05 | |
| F Critical one-tail | 1.300182044 | |

| | *Environment* | *H* |
|---|---|---|
| Mean | 2.50314465 | 0.719811 |
| Variance | 1.26423055 | 0.000798 |
| Observations | 159 | 159 |
| df | 158 | 158 |
| F | 1584.11891 | |
| P(F<=f) one-tail | 1.728E-207 | |
| F Critical one-tail | 1.30018204 | |

| | *Environment* | *D* |
|---|---|---|
| Mean | 2.5031447 | 1.436101 |
| Variance | 1.2642306 | 0.012073 |
| Observations | 159 | 159 |
| df | 158 | 158 |
| F | 104.71287 | |
| P(F<=f) one-tail | 6.91E-115 | |
| F Critical one-tail | 1.300182 | |

| | *Traffic Condition* | *S* |
|---|---|---|
| Mean | 2.496855346 | 17.29931 |
| Variance | 1.264230555 | 0.652137 |
| Observations | 159 | 159 |
| df | 158 | 158 |
| F | 1.938596991 | |

| | *Traffic Condition* | *H* |
|---|---|---|
| Mean | 2.49685535 | 0.719811 |
| Variance | 1.26423055 | 0.000798 |
| Observations | 159 | 159 |
| df | 158 | 158 |
| F | 1584.11891 | |

| | *Traffic Condition* | *D* |
|---|---|---|
| Mean | 2.4968553 | 1.436101 |
| Variance | 1.2642306 | 0.012073 |
| Observations | 159 | 159 |
| df | 158 | 158 |
| F | 104.71287 | |

| | |
|---|---|
| P(F<=f) one-tail | 1.92043E-05 |
| F Critical one-tail | 1.300182044 |
| P(F<=f) one-tail | 1.728E-207 |
| F Critical one-tail | 1.30018204 |
| P(F<=f) one-tail | 6.91E-115 |
| F Critical one-tail | 1.300182 |